\documentclass[preprint,12pt]{elsarticle}

\usepackage{subfig}
\usepackage{lineno,hyperref}
\usepackage{float}
\usepackage{fullpage}
\usepackage{lineno}
\usepackage{amsmath}

\usepackage{amssymb}
\biboptions{numbers,sort&compress}
\usepackage{subfig}
\usepackage{soul}

\def\bfs#1{\mbox{\boldmath{$ #1 $}}}

\usepackage{xcolor}
\usepackage{color}
\definecolor{fgwhite}{rgb}{1,1,1}     
\definecolor{fgred}{rgb}{0.8,0,0}     
\definecolor{Henna}{rgb}{0.63,0.63,0.08}     
\definecolor{fggreen}{rgb}{0,0.5,0}     
\definecolor{fgpurple}{rgb}{0.5,0,1}     
\definecolor{fggray}{rgb}{0.6,0.6,0.7}     
\definecolor{bggreen}{rgb}{0.8,1,0.8}     
\definecolor{fgblue}{rgb}{0,0,0.7}     
\definecolor{bgblue}{rgb}{0.9,0.9,1}     
\definecolor{fgclay}{rgb}{0.51,0.25,0.04}     
\definecolor{paleyellow}{rgb}{0.9,0.9,0.06}

\usepackage{etoolbox}

\journal{---------}

\begin{document}

\begin{frontmatter}



\title{A Fully Discrete Element Approach\\
for Modeling Vacuum Packed Particle Dampers}



\author[1]{Pawel Chodkiewicz}
\author[1]{Robert Zalewski}
\author[2]{Jakub Lengiewicz}

\affiliation[1]{organization={Warsaw University of Technology},
            addressline={Narbutta 84}, 
            city={Warsaw},
            postcode={02-524}, 
            country={Poland}}

\affiliation[2]{organization={Institute of Fundamental Technological Research, Polish Academy of Sciences},
            addressline={Pawinskiego 5B}, 
            city={Warsaw},
            postcode={02-106}, 
            country={Poland}}
            

\begin{abstract}
In this work, we investigate Vacuum-Packed Particle (VPP) dampers -- granular-core dampers offering tunable damping performance -- as a more sustainable alternative to conventional systems such as magnetorheological fluid dampers. A comprehensive computational model of the entire VPP damper system is developed using the Discrete Element Method (DEM). A novel discrete-element model of the flexible foil, responsible for consistently transmitting the external pressure resulting from vacuum application, is introduced and implemented by extending the open-source Yade DEM framework. A prototype VPP damper is also designed and experimentally tested, enabling both model calibration and validation of the simulation results. The calibrated DEM model is subsequently employed in a parametric study to assess the influence of material, geometrical, and process parameters on damper performance. All code, along with the associated experimental and simulation datasets, is made available in an open-access repository.
\end{abstract}



\begin{keyword}
Discrete element method \sep simulations \sep granular damper \sep VPP \sep vacuum-packed particles \sep YADE DEM



\end{keyword}

\end{frontmatter}


\section{Introduction}

In recent decades, growing awareness of environmental issues and the urgent need to comply with evolving ecological standards have fueled the pursuit of more sustainable solutions across various industrial sectors. Researchers and engineers alike are increasingly seeking innovations that integrate eco-friendly “smart” materials capable of adapting their properties in response to changing conditions. Such efforts aim to reduce harmful emissions, promote recycling processes, and ultimately revolutionize engineering practices for a greener and more efficient future \citep{Schwab2017-sw, ELFALEH2023}.

Within this broad class of adaptive and eco-conscious technologies, particulate or granulate materials have garnered significant interest. Granular media can exhibit remarkable behaviors~\citep{Kumar2016, Behringer2018, blanc2021characterization,brigido2019switchable,li2019flexible} such as jamming transitions, enhanced energy dissipation~\cite{Gagnon2019}, and tunable mechanical properties. These characteristics enable their use in a wide range of applications, including blast-mitigation seats in military vehicles \citep{rodak2022possibilities}, medical devices for endoscopic support \citep{loeve2010scopes, loeve2010vacuum}, haptic rehabilitation tools \citep{pinto2019hand}, adaptive sound-absorbing systems \citep{robert2015new}, reconfigurable crash energy absorbers \citep{bartkowski2018concept,bartkowski2019phd}, and flexible grippers in soft robotics \citep{brown2010universal,jiang2012design,kim2013soft}. Moreover, the potential to incorporate recycled or natural fillers into granular cores places these materials squarely within the sustainable product life cycle.

This study focuses specifically on VPP dampers~\citep{ZALEWSKI2016, Rodak2025} -- granular-material-based systems of tunable mechanical damping properties. By leveraging the jamming effect under vacuum, these dampers can transition between quasi-fluid and viscoplastic states, providing a controllable means of vibration damping and impact energy absorption. Moreover, VPP dampers have a relatively simple structure and can be filled with a wide array of granulate media, including those derived from recycled or naturally sourced materials \citep{zalewskiHab}. Such versatility makes VPP dampers a promising replacement for traditional to magnetorheological (MR) or electrorheological (ER) damper devices in applications where simplicity, cost-effectiveness, and environmental compatibility are critical \citep{zalewski2014gubanov, zalewski2014application}.

Characterizing and optimizing granular systems demand advanced mathematical, numerical, and experimental methods. Granular media are inherently complex due to their sensitivity to factors such as humidity, temperature, strain rate, and the interplay of various physical fields. While commercial simulation packages (e.g., LS-Dyna~\citep{benson2009history} or SolidWorks~\citep{Planchard2022SolidWorks}) offer robust tools for macroscopic materials, they are typically less adept at capturing the intricate particle-scale interactions central to granular mechanics. Engineers require specialized methods to accurately capture phenomena such as friction, inter-particle contact forces, and jamming transitions. Thus, precise calibration against experimental data and specialized modeling tools are often indispensable when developing novel devices based on granular media.

The Discrete Element Method (DEM)~\citep{CUNDALL_DEM_1988,CUNDALL_DEM_1992} stands out as a particularly suitable modeling framework for studying particulate systems, given its ability to treat each particle as a separate entity. DEM has been successfully employed in diverse applications: simulating filtration processes with dry and wet approaches \citep{abdallah2024filtration, abdallah2023filtration}, analyzing fluid flow in pre-cracked rock specimens under high pressure \citep{abdi2023simulations}, and modeling electrical conduction in partially sintered porous materials \citep{nisar2024discrete}. It has also proven effective in investigating the damping performance of layered beams with a granular core \citep{bajkowski2015damping}, developing distributed modular robotic systems with reconfigurable mechanical properties \citep{Holobut2014, lengiewicz_podnosniki_PM, piranda2021distributed}, and analyzing extreme loading scenarios in materials such as concrete \citep{benniou2021discrete}. Furthermore, DEM-based studies have been conducted to evaluate the micro-macro behavior of geomaterials under traffic, wave, and earthquake loads \citep{cui2024macro,khakpour2023macro,song2024micromechanical}. 

In our work, we consider VPP damper systems, which rely on interaction between the discrete phase (granulate) and the continuous phase (foil envelope). For studying such more complex scenarios, usually specialized coupling libraries, such as preCICE~\citep{Chourdakis2022}, are utilized to combine dedicated simulation environments for each of the involved phases, see, e.g., fluid-particle-structure interactions~\citep{Adhav2024, AdhavPhD2024}. In the present work, instead, we model the entire foil–granulate system within a single simulation environment. In particular, we develop an extended DEM model that discretizes the elastic foil within the DEM framework, and allows to model its response to mechanical loads and surface pressure loads. This approach follows our earlier first attempts~\citep{chodkiewicz2018discrete}, and provides a robust alternative that demonstrated very good predictive capabilities. We implement the proposed novel DEM-based approach in the open-source Yade~DEM framework~\citep{KOZICKI2008, YadeDEM2021-Zenodo}.

The present work focuses on both the experimental and numerical investigation of Vacuum Packed Particle (VPP) dampers under a variety of operating conditions. The new DEM tool is used to fit experimental data and to examine how various design parameters -- such as particle stiffness, frictional properties, and vacuum level -- affect the overall damping characteristics of VPP systems. As such, our new modeling approach complements prior analytical and semi-empirical studies that employed constitutive equations calibrated with experimental data and advanced algorithmic methods \citep{zalewski2014gubanov, zalewski2014application,mackojc2022preliminary,Rodak2025}.

The remainder of this paper is organized as follows. Section~\ref{sec:experimental_setup} details the experimental setup and procedures used to characterize the mechanical response of the VPP dampers. Section~\ref{sec:modelling} introduces the DEM-based numerical model, describing the contact laws, boundary conditions, and calibration strategies. Section~\ref{sec: model calibration} presents a comparative analysis of the experimental and numerical results, examines the effectiveness of the proposed modeling approach, and provides a parametric study of various VPP damper settings. Finally, Section~\ref{sec: conclusions} concludes the paper with a summary of the main insights and potential future directions.

\section{VPP damper experiment}
\label{sec:experimental_setup}

\begin{figure}[ht!]
	\centering
    \subfloat[]{\includegraphics[height=0.35\textheight]{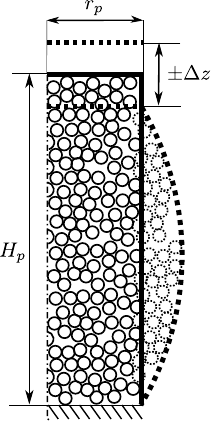}}
          \hspace{0.15\textwidth}
    \subfloat[]{\includegraphics[height=0.35\textheight]{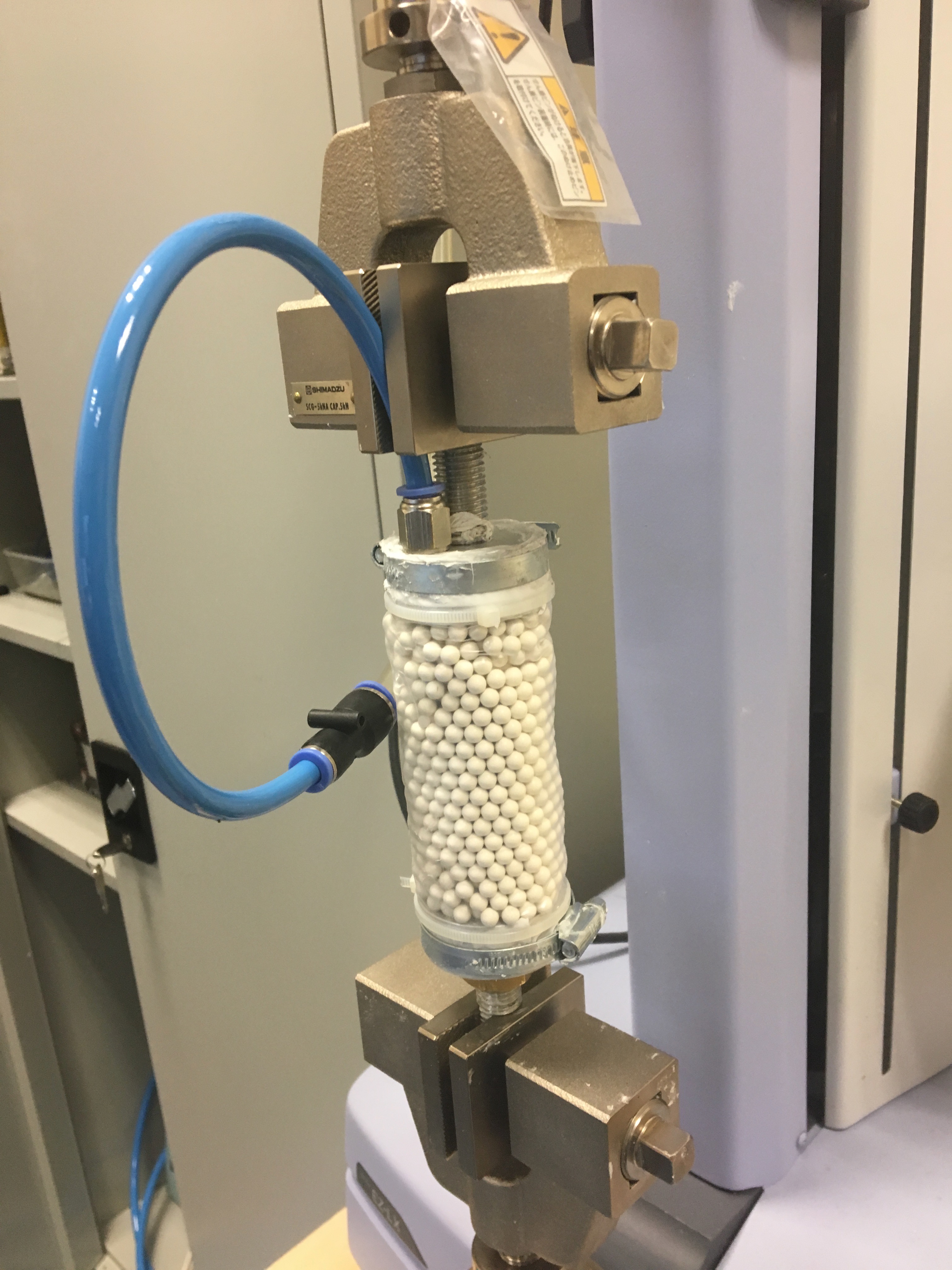}}
        \caption{VPP sample testing stand. (a) schematics of the VPP damper with the dashed line indicating the compressed geometry, (b) the VPP damper fixed in the strength testing machine.}
	\label{fig:probka_VPP}
\end{figure}

A Vacuum Packed Particle (VPP) damper, in its simplest form without any internal spring, consists of a flexible foil shell and loose particles sealed under reduced air pressure. The foil, which is an airtight and durable polymer film, is typically shaped into a cylindrical envelope and filled with a granular medium, see Fig.~\ref{fig:probka_VPP}. In this study, spherical granules are used for that purpose, although virtually any particle type or shape can be employed \citep{zalewskiHab}. Once the granulate is in place, the air inside the envelope is partially evacuated to create a pressure difference. This reduction in internal pressure causes the particles to compact and lock together through frictional and interlocking forces, effectively transitioning the material from a fluid-like state to a quasi-solid or rigid state. Consequently, the damper exhibits enhanced stiffness and damping characteristics under vacuum, as the granules are no longer free to move independently when subjected to external mechanical loading. Furthermore, the level of underpressure can be controlled, allowing to modulate the energy dissipation performance under oscillatory motion.

In the experiment, we use a VPP damper with a height of $H_{\text{p}}=100$~mm and a radius of $r_{\text{p}}=25.5$~mm, and the foil thickness of $t_{\text{f}}=0.15$~mm. The geometry of the setup is presented in Fig.~\ref{fig:probka_VPP} and the list of callibrated parameters is provided in Table~\ref{tab:calibrated parameters}). The experimental procedure consists of the following steps:
\begin{enumerate}
    \item \textbf{Filling the envelope:} The foil envelope is filled with spherical granules.
    \item \textbf{Compression and re-filling:} The granulate is iteratively compressed and re-filled to achieve a higher packing density. (The volume ratio was measured \textit{a~posteriori}, after the experiment concluded.) 
    \item \textbf{Sealing the sample:} The sample is capped and sealed at both ends, with the upper cap integrated into a compressor that controls the underpressure level within the sample.
    \item \textbf{Mounting in the Testing Machine:} The sample is placed in the grips of the SHIMADZU EZ-LX testing machine equipped with a 5000 kN force load cell.
    \item \textbf{Pressure Stabilization:} The underpressure is set to one of four distinct values, $\Delta p$, of 0.02~MPa, 0.04~MPa, 0.06~MPa, and 0.08~MPa, corresponding to 20\%, 40\%, 60\%, and 80\% of atmospheric pressure, respectively.
     \item \textbf{Cyclic Loading:} Under a constant quasi-static strain rate and a triangular excitation protocol, the sample is subjected to cyclic compression and tension over a displacement amplitude of $\Delta z=10$~mm. Throughout this process, the underpressure level is maintained, and time, yaw displacement, and force data are recorded throughout the cycles. (The initial force is calibrated to 0~N.)
\end{enumerate}

\begin{figure}[h]
	\centering
          \includegraphics[width=0.5\textwidth]{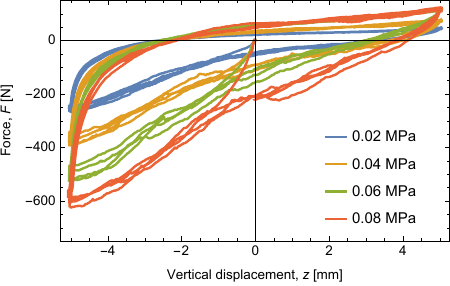}
        \caption{Experimental results for four different underpressure levels $\Delta p$, which represents the pressure difference between the atmospheric pressure and the pressure inside the damper. The hysteresis loops for the first four cycles are shown for each case.}
	\label{fig:wyniki_eksp}
\end{figure}

The results of the experiment are shown in Fig. \ref{fig:wyniki_eksp}. We can observe a significant effect of the underpressure level on energy dissipation, with higher underpressure yielding larger hysteresis loops. This effect allows for tuning of the system: by adjusting the partial vacuum within the device, one can control its overall mechanical properties. Another observation is that, after the short initial phase, the process stabilizes, and the system maintains consistent energy dissipation characteristics over subsequent cycles. These observations confirm the robustness of VPP dampers, and underscore its potential for applications requiring adaptable damping characteristics. 

\section{DEM Modeling} 
\label{sec:modelling}

In the presented approach, the mechanical response of the VPP damper system is modeled using an explicit Discrete Element Method (DEM) framework implemented in Yade~\citep{KOZICKI2008, YadeDEM2021-Zenodo}. This section briefly summarizes the essential DEM operations, see Section~\ref{sec: DEM simulation framework}, and then describes our novel approach to representing the flexible foil as a two-dimensional network of spherical particles, see Sections~\ref{sec: DEM model of foil} and~\ref{sec: DEM model of foil vacuum}. All code, together with the experimental and simulation datasets, is made available in open-access repositories~\cite{chodkiewiczGitHub2025vpp, chodkiewiczZenodo2025vpp}.

\subsection{DEM Simulation Framework}
\label{sec: DEM simulation framework}

In our DEM model of VPP damper, we distinguish tree types of entities: the particles of granulate, the particles used to discretize foil, and the rigid planes/segments to model the upper and lower capping of the VPP sample. For the interactions between the granulate particles we use the Coulomb friction law. The interaction between the particles of foil are of adhesive type, allowing for modeling permanent bonds that represent the elastic response of foil (see also Sec.~\ref{sec: DEM model of foil} and Sec.~\ref{sec: model calibration}). The third type of interaction is between the granulate and the walls (i.e., the foil and the upper/lower capping). In real system, these interactions involve friction, however, in the current DEM model, we decided to model the granulate-foil interaction as frictionless. The reason for that is that, in our approach, the originally smooth foil is discretized as a "rough" layer of DEM particles, therefore, there exist a dilatant interlocking between the granulate and foil particles. This granulate-foil interlocking provokes the friction-like resistance in sliding, even when no granulate-foil friction is present.

After the VPP damper DEM geometry is established, the multi-step DEM analysis is performed. In the analysis, at a given time step, the following sequence of actions is performed:
\begin{enumerate}
    \item \textbf{Collision detection:} Identify when two particles come into potential contact based on their positions and radii.
    \item \textbf{Interaction initialization and property assignment:} Create the interaction and determine its properties, such as stiffness. This is done either through precomputed values or derived from the physical properties of the interacting particles. 
    \item \textbf{Interaction physics:} For interactions already established in previous time steps, perform the following computations:
    \begin{enumerate}
        \item \textbf{Strain evaluation:} Calculate the deformation of the interaction based on relative particle displacement.
        \item \textbf{Force computation:} Determine forces as functions of the calculated strains and external loadings, using material-specific constitutive models. The particular example of external loading due to the pressure difference is presented in Sec.~\ref{sec: DEM model of foil vacuum}.
    \end{enumerate}
    \item \textbf{Force application:} Apply the computed forces to the particles involved in the interaction to update their motion and positions. Increment the time, apply the Dirichlet boundary conditions, and proceed to the Step~1.
\end{enumerate}
This sequence is repeated until the simulated process finishes, e.g., a desired number of loading cycles is performed.


\subsection{DEM Modelling of Flexible Foil: elastic model}
\label{sec: DEM model of foil}

\begin{figure}[t!]
	\centering
    \subfloat[]{
	\includegraphics[width=0.3\textwidth]{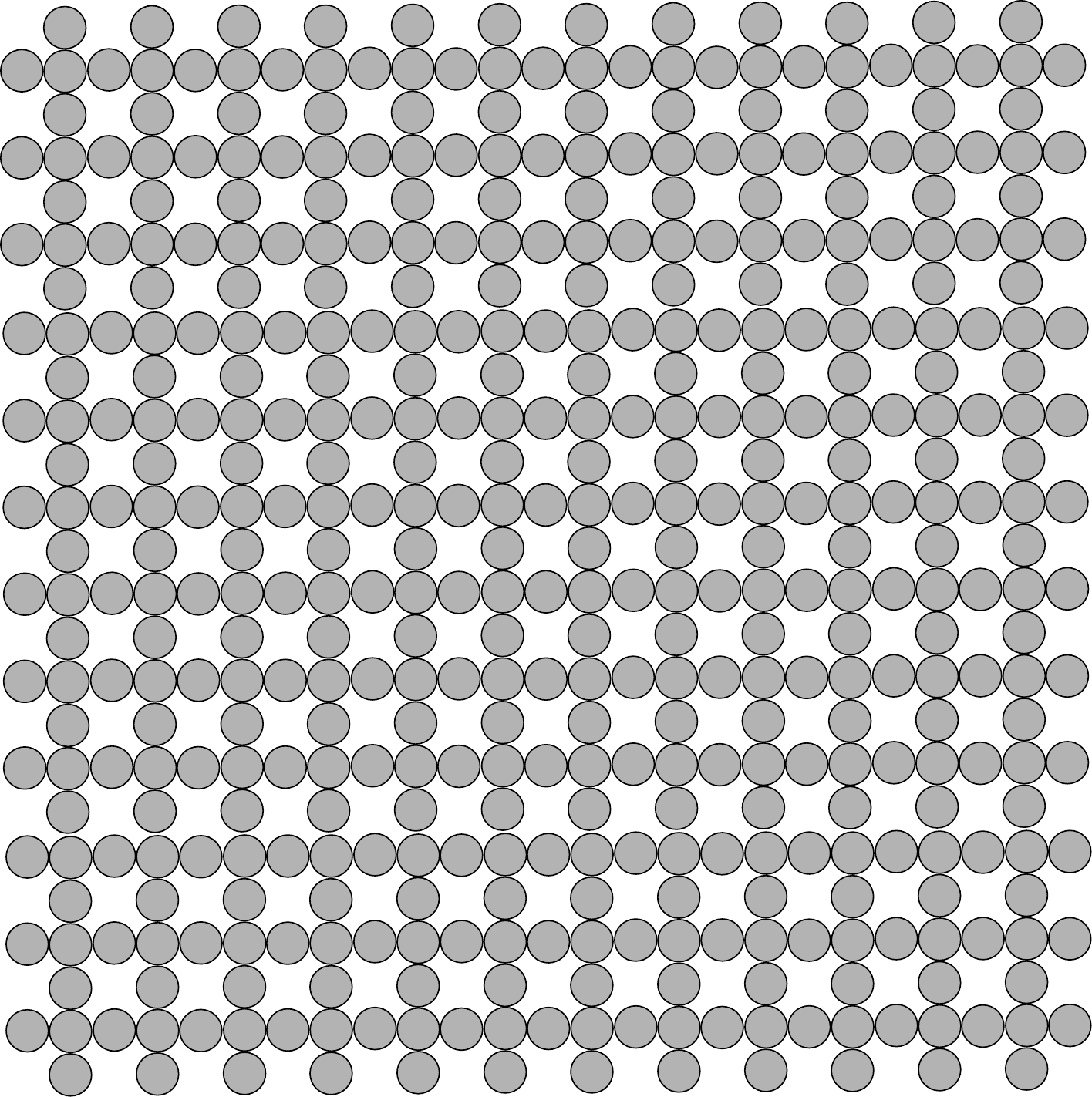}
    }
    \hspace{0.03\textwidth}
    \subfloat[]{
        \includegraphics[width=0.3\textwidth]{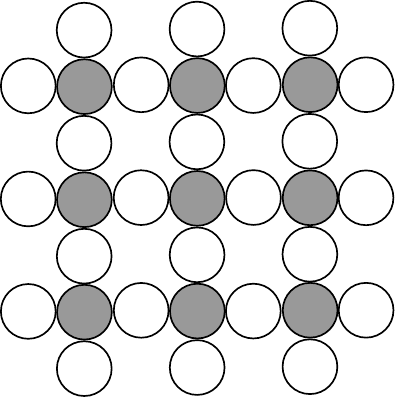}
    }
    \hspace{0.03\textwidth}
    \subfloat[]{
        \raisebox{0.05\textwidth}{\includegraphics[width=0.2\textwidth]{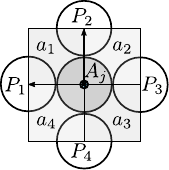}}
    }
	\caption{Schematic of the arrangement of particles in the membrane (a). In the zoom-in (b), the particles used to apply pressure difference are marked in gray color. In (c), the elementary cell $j$ is shown, with the notation used for calculating the surface area vector $\bfs{A}_j$.}
		\label{fig:yade_folia_cala}
\end{figure}

We model the flexible foil membrane as a two-dimensional network of interconnected spherical particles. There are countless possibilities in which circular particles can be arranged in two-dimensional space~\citep{williams1979circle, blokhuis1989sphere, steinhaus1999mathematical}, with square- and hexagonal packings being the most popular choices. When making that choice in our particular case, we were aiming to achieve a lightweight DEM representation of foil that will assure robust and accurate predictions at a minimum additional computational burden. This led us to the square-based arrangement of foil particles that contains regular vacancies, see Fig.~\ref{fig:yade_folia_cala}a. 

This special particle arrangement serves for several purposes. Firstly, the choice of the square packing is less dense as compared to the hexagonal packing, which is additionally optimized by removing selected 25\% of particles, allowing for a significant reduction in computation time. Secondly, the removal of some particles provides more flexibility to the foil membrane, which is now more capable to adapt effectively to the shape of the granules within the sample, which allows to optimally transmit forces associated with the pressure difference. Lastly, the remaining network of particles that form the foil is able to accurately represent the elastic properties of foil and to transmit and re-distribute forces associated with the cyclic loading of VPP damper.

The orthogonal orientation of the particle network in the cylindrical VPP envelope is chosen such that the main symmetry axis coincides with the longitudinal and circumferential directions of the cylinder, see also Fig.~\ref{fig:yade_folia_cala2}. These are the two main directions that transmit the tensile forces in the discretized foil. In order to create the relationship between the elastic modulus of real (continuous) foil, $E_{\text{y}}$, and the elastic modulus of its simplified discretized model, $E_{\text{s}}$, we need to consider two effects. The first effect is related to the fact that there are vacancies in the discretized model, and only every second column/row can transmit the forces. This makes it necessary to effectively increase the elastic modulus of the discretized model by the factor of 2 with respect to the continuum counterpart. The second effect is related to difference between the thickness of the real foil, $t_{\text{f}}$, and the apparent thickness of the discretized model which is the diameter of the foil particle, $2 \cdot {r_{\text{s}}}$. The elastic modulus must be accordingly scaled to account for that. The final formula for the corrected elastic modulus reads
\begin{equation}
{E_{\text{s}}} =2 \cdot {E_{\text{y}}} \cdot \frac{t_{\text{f}}}{2 \cdot {r_{\text{s}}}}.
\label{eq: mod_younga_kul2}
\end{equation}

\subsection{DEM Modelling of Flexible Foil: vacuum pressure}
\label{sec: DEM model of foil vacuum}

As explained in Section~\ref{sec:experimental_setup}, the operation of Vacuum Packed Particle dampers is controlled by the level of vacuum inside the envelope. This results in external pressure that is transmitted to the granulate through the foil envelope. In the DEM simulation, this external loading can be applied as forces acting at foil particles in the direction perpendicular to the current surface of the membrane.  

In the case of foil discretization proposed in this work, the forces are only applied to the foil particles that are at the intersection of perpendicular chains forming the network, see gray particles in Fig.~\ref{fig:yade_folia_cala}b. This choice is motivated by the fact that, for the particles at the intersection, it is straightforward to determine the tributary area vector that is necessary to produce the force vector. Even though this heuristic is only an approximation, it proved to be sufficient to capture the pressure dependence of VPP damper responses, as shown later in this work.

For the proposed regular geometry, the tributary surface area vector $\bfs{A}_j$ of the foil particle $\bfs{S}_j$ is the sum of respective contributions from four neighboring segments:
\begin{equation}
\label{eq: A_j surface area vector}
{{\bfs{A}}_j} = \sum_{i=1}^{4}{{\bfs{a}}_{i}}, \qquad\qquad {{\bfs{a}}_{i}} = ({{\bfs{P}}_i} - {{\bfs{S}}_j}) \times ({{\bfs{P}}_{1 + (i\% 4)}} - {{\bfs{S}}_j})
\end{equation}
where $\bfs{S}_j$ is the position of the central (gray) particle, $\%$ is the modulo operation, and $\bfs{P}_i$ is the position of the $i$-th neighboring particle, see Fig.~\ref{fig:yade_folia_cala}c. $\bfs{A}_j$ is pointing outwards to the membrane surface, and depends on the current position of the contributing particles. For such a defined tributary area vector, the force applied on a foil particle $\bfs{S}_j$ is equal to
\begin{equation}
\bfs{F}_j = -\Delta{}p \bfs{A}_j,
\end{equation}
where $\Delta{}p$ is the pressure difference (positive for the underpressure). 



\section{Model calibration, performance and parametric study}
\label{sec: model calibration}

This section provides an empirical demonstration of the fitting and predictive capabilities of the proposed DEM model. In Section~\ref{sec: setting up DEM analysis}, we outline the steps to set up the Yade DEM simulation to represent the experimental setup. Section~\ref{sec: calibration of VPP damper model} details the calibration of the DEM model parameters using experimental data. Finally, Section~\ref{sec: parametric study} validates the predictive performance of the calibrated model, and examines the influence of material, design, and process parameters on VPP damper performance.


\subsection{DEM analysis for VPP damper}
\label{sec: setting up DEM analysis}

\begin{figure}[ht]
	\centering
         \subfloat[]{
	\raisebox{0.03\textwidth}{\includegraphics[width=0.4\textwidth]{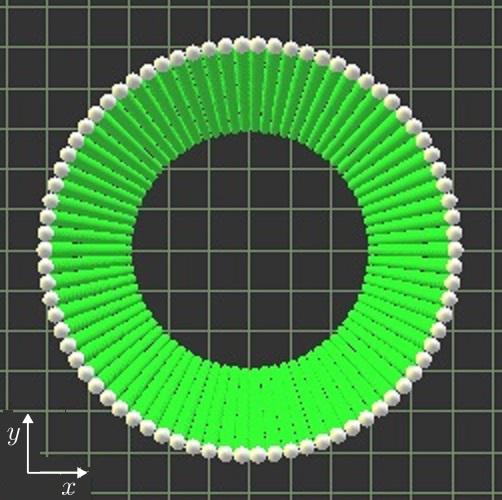}}
    }\hspace{0.02\textwidth}
     \subfloat[]{
	\includegraphics[width=0.5\textwidth]{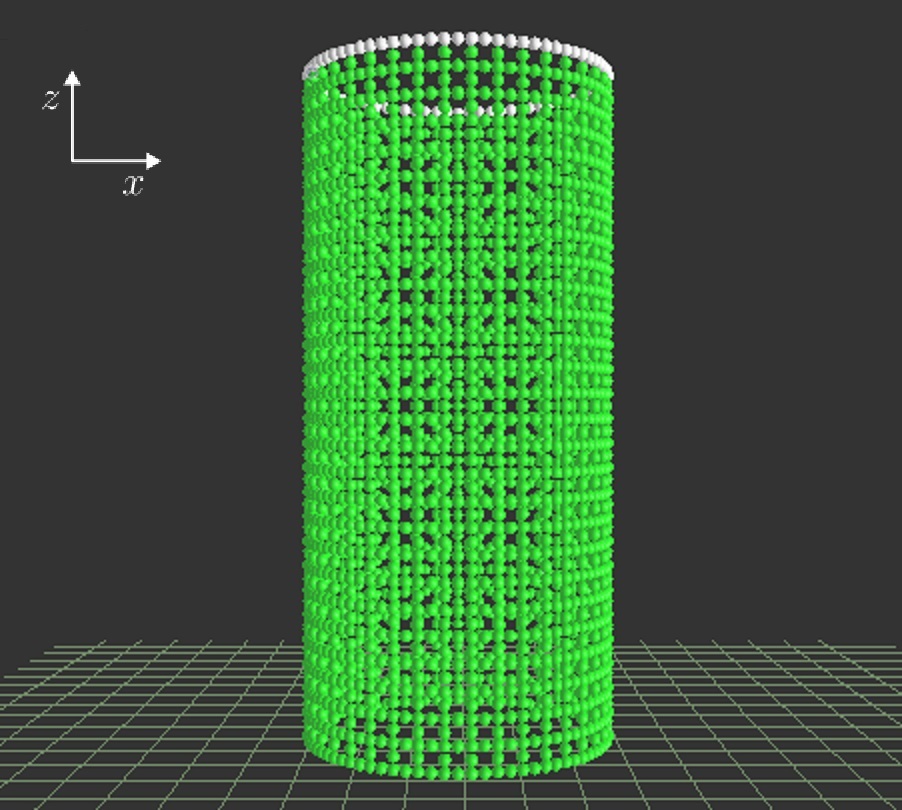}
    }
	\caption{The foil membrane geometry. Snapshot from the Yade DEM simulator.}
		\label{fig:yade_folia_cala2}
\end{figure}

In the DEM analysis, we aim to accurately simulate the laboratory tests of the VPP damper described in Section \ref{sec:experimental_setup}. We begin with setting up the foil membrane by arranging the particles to form the cylindrical shape, as illustrated in Figure \ref{fig:yade_folia_cala2}, following the arrangement topology discussed in Section~\ref{sec: DEM model of foil}. The number of particles in the circumferential direction is adjusted such that the radius of the DEM model is as close as possible to the real radius, $r_{\text{p}}$, used in the experiment. The upper and lower layers of the foil are fixed, which allows us to control their positions during the simulation.

\begin{figure}[h!]
	\centering
        \subfloat[]{
            \includegraphics[width=0.49\textwidth]{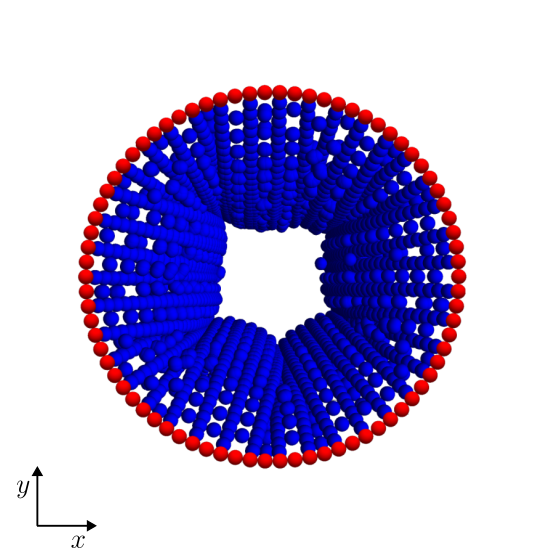}
        }
        \subfloat[]{
               \includegraphics[width=0.49\textwidth]{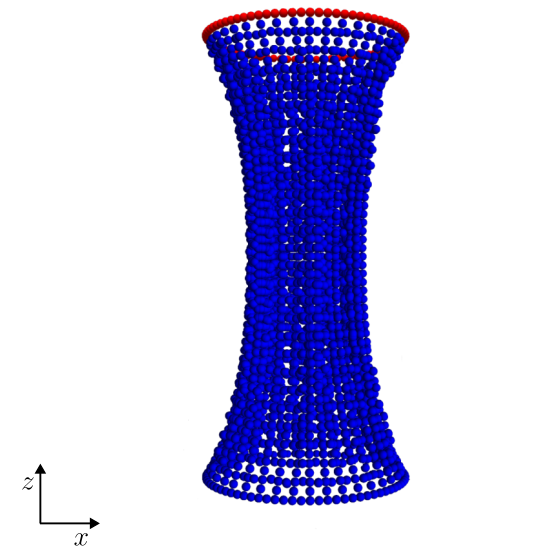}
        }
        \caption{Simulation of an empty foil envelope subjected to an underpressure corresponding to the~10\% reduction with respect to the atmospheric pressure ($\Delta p=0.01$~MPa).}
	\label{fig:podcisnie}
\end{figure}

Having the membrane geometry ready, we perform a sanity check of the proposed novel mechanism of applying pressure difference (see also Section~\ref{sec: DEM model of foil vacuum} for the details of DEM formulation). In this simple qualitative test, we used an empty foil envelope and applied various levels of pressure or underpressure. The test confirmed the expected behavior of the model. For instance, the result for a moderate level of underpressure is presented in Fig.~\ref{fig:podcisnie}, showing a squeezed deformation pattern with several visible buckling folds, which also involves the self-contact of the foil folds.

\begin{figure}[h!]
	\centering
     \subfloat[]{
	\includegraphics[width=0.3\textheight]{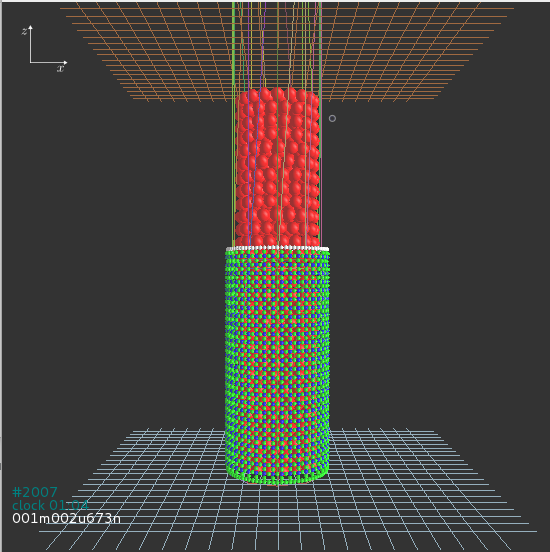}
        }\hspace{50pt}
         \subfloat[]{
	\includegraphics[width=0.3\textheight]{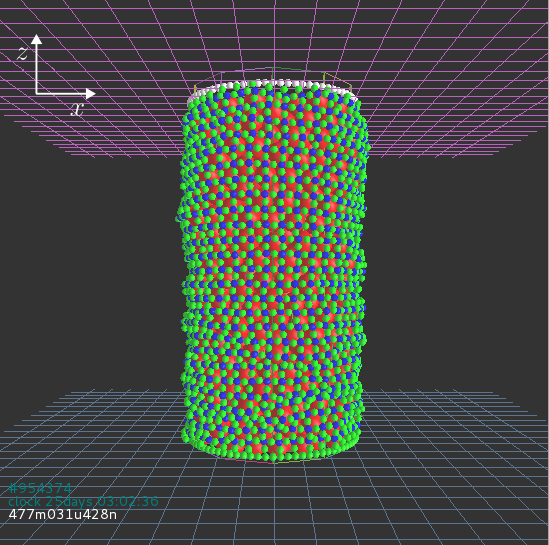}
    }
	\caption{Two stages of the precess of filling the foil with the granulate, with the packing level of 54\%. a)~The initial stage, in which the predefined number of particles is generated and placed inside the foil cylinder and in the auxiliary rigid "funnel" above the cylinder. b)~The filled sample, after allowing the granulate to fall freely under gravity and then compressing them by the upper wall.
}
	\label{fig:filling_the_granulate}
\end{figure}

In the next step, the empty membrane is filled with the granulate by letting the granules naturally fall under the gravity forces. To ensure complete filling of the sample to a desired packing ratio of 54\%, the necessary amount of granulate is created and directed into the membrane through a special temporary "funnel" structure that is constructed with triangular rigid facets. 
During the filling process, the friction between the granulates is reduced to zero to facilitate their free movement, and consequently allowing for a non-obstructed filling of the entire foil structure. As the last step, a rigid capping is moved downwards until a designed damper's height, $H_\text{p}$, to additionally enforce the granulates' rearrangement.
This process is visualized in Figure~\ref{fig:filling_the_granulate}.

The main loading cycles in the simulation are carried out by applying a sawtooth-shaped vertical displacement with an amplitude of \( \Delta z = 10\,\text{mm} \) to the upper wall and the top ring of foil particles. The loading velocity is set to \( v_\text{z} = 33.3 \times 10^{-5}\,\text{m/s} \) (see also the parametric study in Section~\ref{sec: parametric study} and Fig.~\ref{fig:velocity}). The vertical reaction forces are captured individually at the upper wall and at the fixed top foil ring, which allows their individual contributions to be observed; see also Fig.~\ref{fig: model calibration}. The forces are sampled 500 times per cycle (i.e., every \( 0.12~\text{s} \)) by averaging over the interval \( 2 \times 10^{-4}~\text{s} \), where \( t_\text{c} = 2 \Delta z / v_\text{z} \simeq 60~\text{s} \) is the duration of a single loading cycle, while the time increment in the DEM analysis is set to approximately \( 2 \times 10^{-6}\,\text{s} \).

\subsection{Calibration of the parameters of VPP damper model}
\label{sec: calibration of VPP damper model}


The DEM model of the VPP damper includes several material and geometrical parameters, accompanied by the respective process parameters, as summarized in Table~\ref{tab:calibrated parameters}. Most of these parameters were obtained through direct measurements. Regarding the granulate, the granulate radius $r_{\text{k}}$ and density $\rho_{\text{k}}$ represent average values calculated from measurements performed on a sample of 100 particles. The Young modulus of the granulate was taken from the manufacturer's specification. This approach is justified because, as demonstrated by the parametric study presented in Fig.~\ref{fig:young}, the response of the VPP damper remains insensitive to variations in the Young modulus across a wide range of values.

\begin{table}[h!]
\centering
\begin{tabular}{|l|c|c|c|}
\hline
Parameter & Symbol & Value & Unit \\ \hline
Granulate: particle radius  &       $r_{\text{k}}$    & $2.965 \times 10^{-3}$  & m         \\ \hline
Granulate: particle density  &        $\rho_{\text{k}}$        &2689.639 &kg/m$^3$   \\ \hline
Granulate: particle Young modulus & $E_{\text{k}}$ & 2300&MPa \\ \hline
Granulate: particle friction coefficient & $\mu_{\text{k}}$ & 0.6 & - \\ \hline
Granulate: volume ratio of grains &&54\%  & - \\[0.4em] \hline
Membrane: foil thickness & $t_{\text{f}}$ & $0.15 \times 10^{-3}$  & m         \\ \hline
Membrane: particle radius &       $r_{\text{f}}$    & $r_{\text{k}}/3$  & m         \\ \hline
Membrane: material density &        $\rho_{\text{f}}$        & 151.77 &kg/m$^3$   \\ \hline
Membrane: particle Young modulus (Eq.~\eqref{eq: mod_younga_kul2})  & $E_{\text{s}}$ & 18.545
& MPa \\ \hline
Membrane: real foil Young modulus  & $E_{\text{f}}$ & 122.18 & MPa \\ \hline
Membrane: friction coefficient of particles & $\mu_{\text{f}}$ & 0 & - \\[0.4em] \hline
Process: amplitude of excitation & $\Delta z$ & $10.0 \times 10^{-3}$ & m\\ \hline
Process: loading velocity & $v_\text{z}$ & $33.3 \times 10^{-5}$ & m/s\\ \hline
\end{tabular}
\caption{Calibrated parameters of the VPP damper.}
\label{tab:calibrated parameters}
\end{table}

\begin{figure}[h!]
		\centering
         \subfloat[]{
		\raisebox{2.5em}{\includegraphics[width=0.47\textwidth]{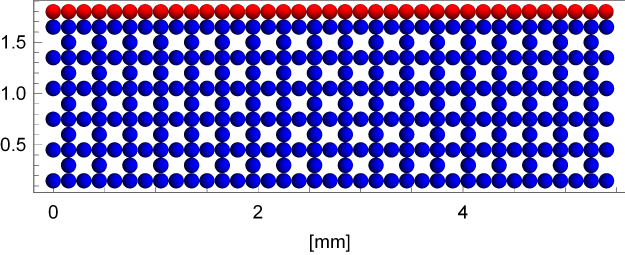}}}
            \hspace{1em}
		 \subfloat[]{
         \includegraphics[width=0.47\textwidth]{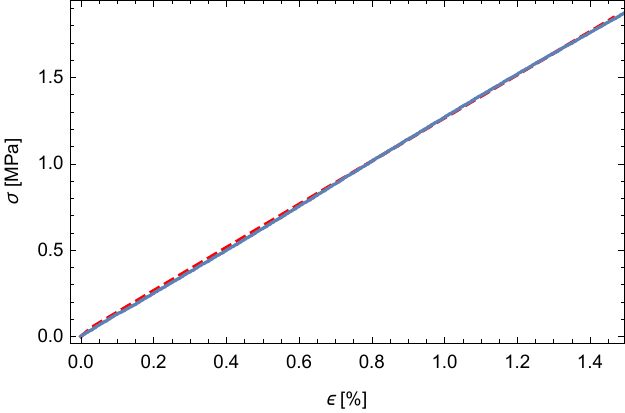}
         }
        \caption{Left: Membrane geometry in Yade DEM simulation (width x height: 37x12 particles).
Right: Uniaxial stretching results of the membrane: Yade DEM model (red dashed) vs. experiment (blue).}
		\label{fig:foilia}
\end{figure}

The foil parameters, such as thickness and density, were determined through direct measurements. The Young modulus of the foil was obtained via an auxiliary dedicated experiment in which a rectangular foil membrane was subjected to a uniaxial tensile test. The experimentally determined elastic modulus was then used to validate the DEM model for the foil according to Equation~\ref{eq: mod_younga_kul2}. The validation results indicate a very good agreement between the experimental data and the model predictions, as shown in Fig.~\ref{fig:foilia}.

The above-mentioned procedure allowed the determination of all model parameters except for the friction coefficient between granulate particles, $\mu_{\text{k}}$. The friction coefficient, when obtained from direct experimental measurement, was at the level of 0.2, which  could not provide a sufficiently precise fit between the VPP damper experiment and the DEM simulation. One of the reasons for that could be geometrical inhomogeneities of the particles' surface, leading to a more pronounced frictional response under the multi-axial load conditions. The problem of fitting the friction coefficient for particle damping systems has also been observed by other researchers~\cite{Gagnon2019}. To alleviate that problem, we calibrated the friction coefficient by fitting the results of the DEM simulation (outlined in Section~\ref{sec: setting up DEM analysis}) to the experimental data shown in Fig.~\ref{fig:wyniki_eksp}.  Calibration was conducted only for the single underpressure level of $\Delta p = 0.06$~MPa, employing a grid search for friction coefficient values $\mu_{\text{k}}$ ranging from $0$ to $1$ (the parametric study for various friction coefficients, which served for the grid search, is shown in Fig.~\ref{fig:friction}).

\begin{figure}[h!]
    \centering
    \includegraphics[width=0.5\textwidth]{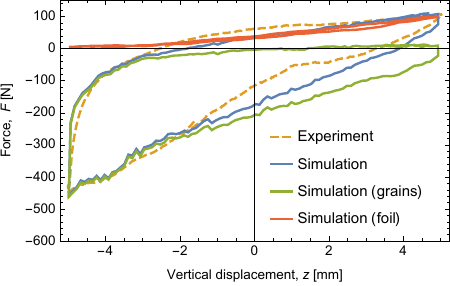}
		\caption{Comparison of the prediction of the calibrated model and the experimental response for $\Delta p=0.06$~MPa. The simulated response force is additionally broken down into the individual contributions from foil and grains (granulate).}
        \label{fig: model calibration}
\end{figure}

The calibration results for $\Delta p = 0.06$~MPa are presented in Fig.~\ref{fig: model calibration}, along with a detailed breakdown for the individual reaction force contributions from foil and granulate. Such decomposition, obtainable only through the DEM simulation, offers deeper insights into the respective roles of these two main components of the VPP damper. The comparison demonstrates very good qualitative and quantitative agreement between simulation and experimental results. Slight discrepancies, observed at the onset of compression phase, may stem from operational limitations of the vacuum pump, which struggled to keep pace with the decreasing sample's volume. 

\begin{figure}[h!]
    \centering
    \subfloat[$\Delta p=0.02$ MPa]{
            \includegraphics[width=0.49\textwidth]{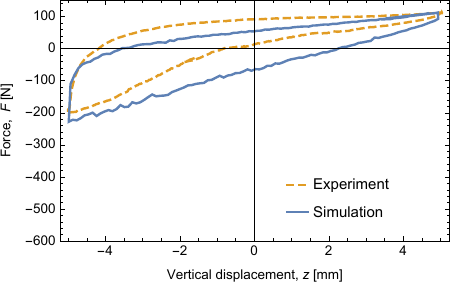}
            }
    \subfloat[$\Delta p=0.04$ MPa]{
            \includegraphics[width=0.49\textwidth]{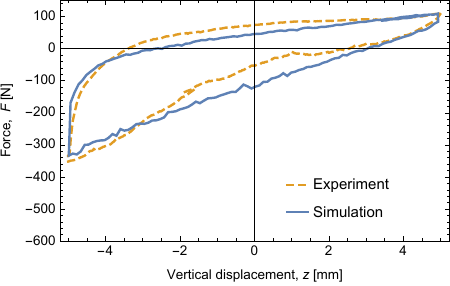}
            }  \\
    \subfloat[$\Delta p=0.06$ MPa]{
            \includegraphics[width=0.49\textwidth]{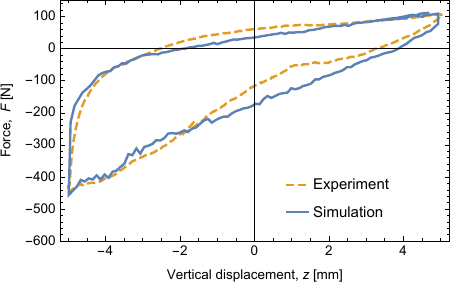}
            }
    \subfloat[$\Delta p=0.08$ MPa]{
            \includegraphics[width=0.49\textwidth]{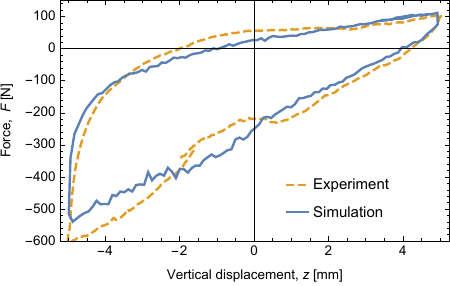}
            }
		\caption{Performance of the calibrated model for different pressures: (a) $\Delta p=0.02$~MPa, (b) $\Delta p=0.04$~MPa, (c) $\Delta p=0.06$~MPa (the case used for calibration), (d) $\Delta p=0.08$~MPa.}
\label{fig: simulation-experiment comparison for underpressures}
\end{figure}

Fig.~\ref{fig: simulation-experiment comparison for underpressures} showcases the model's predictive capabilities for underpressure levels that were not part of the calibration dataset. The maintained accuracy across these cases further confirms the robustness of the calibrated DEM model.




\subsection{Parametric study}
\label{sec: parametric study}

\begin{figure}[t!]
    \centering
    \includegraphics[width=0.5\textwidth]{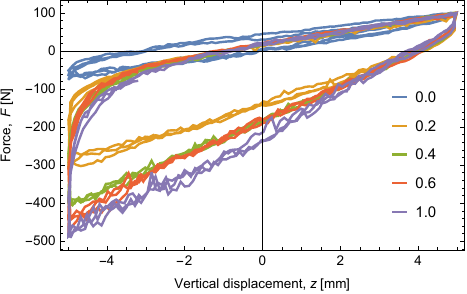}
    \caption{Parametric study of the model response to the granulate friction coefficient $\mu_{\text{k}}$. Vacuum pressure $\Delta p=0.06$~MPa.}
    \label{fig:friction}
\end{figure}

The calibrated DEM model is now used to study the influence of selected material, design, and process parameters on the performance of VPP dampers. The first analyzed parameter is the friction coefficient between the granulate particles—already employed during the calibration of the DEM model; see Section~\ref{sec: calibration of VPP damper model}. As shown in Fig.~\ref{fig:friction}, modulating friction allows control over the hysteresis loop, and consequently, over the dissipative properties of VPP dampers. This effect on dissipation is strongly nonlinear: initially, even a slight increase in the friction coefficient significantly changes the hysteresis loops; however, beyond a certain point, further increases in friction have only a minor impact on dissipation. For the studied case, changing the friction coefficient from 0.6 to 1.0 results in only a few percent difference in the energy dissipation capacity of the VPP damper. Another interesting observation is the dissipative behavior at \( \mu_\text{k}=0 \), which supports the hypothesis that the dynamic reorganization of the jammed particle system is responsible for dissipative effects. In summary, the friction coefficient between particles is an important factor in designing VPP dampers with desired properties.

\begin{figure}[t!]
    \centering
    \includegraphics[width=0.5\textwidth]{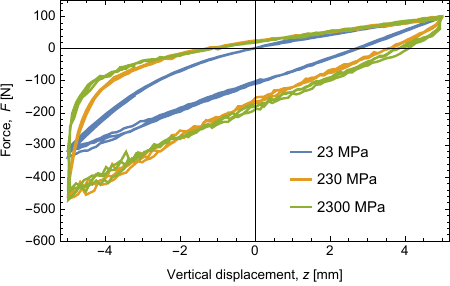}
		\caption{Parametric study of the model response to the Young Modulus of the grains inside the core. Vacuum pressure $\Delta p=0.06$ MPa }
        \label{fig:young}
\end{figure}

The second studied material property of the system is the elastic modulus of particles. Figure~\ref{fig:young} illustrates the impact of Young's modulus on the overall dissipative characteristics of the VPP damper. Similarly to the previously described case of the friction coefficient, the influence of elastic modulus on energy dissipation capability is clearly visible. Again, a nonlinear relationship can be observed. Increasing stiffness tenfold, from 23 to 230 MPa, significantly affects the damping efficiency of the sample; however, this effect saturates beyond a certain threshold. Such behavior is, to some extent, intuitive. For conglomerates made of compliant grains, such as silicones, gels, or soft rubbers, the particles deform considerably under combined multi-axial loading. Consequently, the particle conglomerate tends to deform as a whole rather than reorganize, which effects in reducing damping effects. In contrast, the use of granulates of higher stiffness promotes the rearrangement of grains as the main mechanism of the conglomerate's deformation, resulting in an intensive dissipative response. Note that this observation partially supports the hypothesis that the dynamic rearrangement is the main dissipative mechanism in VPP systems. More direct evidence could be provided by an experimental validation for the case of softer particles. We see this as an interesting avenue for future research.  

\begin{figure}[t!]
    \centering
    \includegraphics[width=0.5\textwidth]{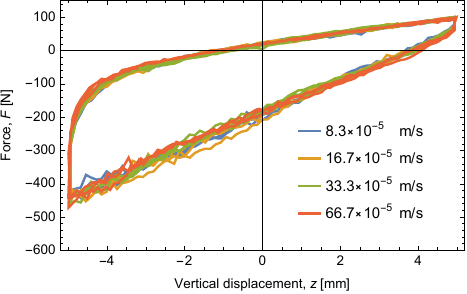}
		\caption{Parametric study of the model response to loading velocity, \( v_\text{z}\). Vacuum pressure $\Delta p=0.06$~MPa}
        \label{fig:velocity}
\end{figure}

Now, we are going to study the effect of the imposed oscillation rate, which is a process parameter that describes the speed of linear displacement of the grips in the uniaxial testing machine. The sensitivity of the simulated VPP damper response to the loading velocity is illustrated in Fig.~\ref{fig:velocity}. According to the DEM model results, the strain rate does not significantly influence the predicted global dissipative properties of the VPP dampers. These results align with expectations since, in the DEM model, both the material properties of the foil and particles are linearly elastic, and the Coulomb friction model employed is rate-independent. Consequently, the numerical model is not expected to directly exhibit any viscous or rate-dependent effects over a wide range of loading velocities. The dominating dissipative mechanism is through the rearrangement of particles, having its roots in the damping of the kinetic energy of particles released from the jammed state. This rate-independent nature of VPP dampers is also partially confirmed by experimental study~\cite{Rodak2025} for different family of VPP dampers, which additionally justifies the modeling assumptions of the present model. However, we expect that for real system, imposing more pronounced oscillation rates can indeed result in differences in the shape of hysteresis loop. This effect can be caused either by the insufficient performance of the vacuum pump (already mentioned in Section~\ref{sec: calibration of VPP damper model}) or by an increased coupling between the global response and the local dissipation of particles' kinetics. A dedicated experiment for such high-velocity deformation cases can be an interesting topic for future study.  

\begin{figure}[t!]
    \centering
    \includegraphics[width=0.5\textwidth]{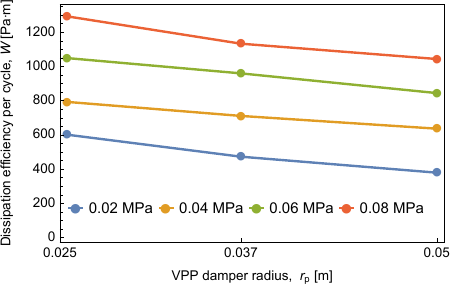}
		\caption{Analysis of the effect of sample diameter on the damping efficiency, $W$, for four different underpressure levels $\Delta p$. The base diameter used for all earlier examples is $r_\text{p} = 0.0255$ m.}
        \label{fig:effect of sample diameter}
\end{figure}

Finally, we demonstrate how changing the geometry of VPP dampers affects their dissipative performance. To illustrate this, we vary the damper radius, $r_{\text{p}}$, and measure the total energy dissipated per cycle, expecting it to increase with radius. To quantify the \emph{efficiency} of energy dissipation per cycle, $W$, the total energy dissipated is normalized by the damper's cross-sectional area, $s_\textrm{s}=\pi r_{\text{p}}^2$. Given a set of force-displacement pairs, $(\Delta z_i, F_{i})$, representing the damper's response over one loading cycle, the dissipation efficiency is defined as:  
\begin{equation}
    W = \frac{1}{2s_{\textrm{s}}}\sum_i^n \left (
    \Delta z_i F_{i+1} -  \Delta z_{i+1} F_{i} 
    \right ),
\end{equation}
where the trapezoidal rule is employed to compute the dissipated energy as the area enclosed by the hysteresis loop. Results for three damper radii and four distinct underpressure levels are presented in Fig.~\ref{fig:effect of sample diameter}. It is evident that increasing the damper's radius at constant height reduces the energy dissipation efficiency, a trend consistent across all analyzed underpressure levels. Consequently, to enhance dissipation performance, stacking multiple slender VPP dampers is preferable to using a single damper with a larger radius.

\section{Conclusions and future work}
\label{sec: conclusions}

In this work, we proposed a novel approach to modeling Vacuum Packed Particles (VPP). In contrast to the commonly adopted phenomenological models, which treat VPP systems as continua, we employed a fully discrete framework. The Discrete Element Method (DEM) was used to model not only the granular core but also the continuous elastic membrane that separates the granulate from the external environment and transmits the pressure difference induced by the applied vacuum. The membrane was represented as a system of interconnected spherical particles, with a dedicated model developed to impose the corresponding surface forces. A dedicated VPP damper prototype was built, and a series of experiments were conducted to calibrate the model parameters and validate the simulation results. The calibrated DEM model was shown to generalize well to underpressure levels not included in the fitting procedure. Furthermore, a parametric study was conducted to analyze the influence of material, process, and geometrical design parameters on the macroscopic response of VPP-based systems, demonstrating the capability of the proposed approach for the tuning and optimization of VPP dampers.

Several directions for future research have been identified. The most immediate continuation would involve a more extensive parametric study, encompassing both DEM simulations and corresponding experimental validation. Such investigations could reveal additional effects not captured by the present model. Another promising direction is a more detailed analysis of the microscopic mechanisms responsible for macroscopic dissipation, in particular the respective contributions of frictional sliding and dynamic particle reorganization. The developed DEM model offers a suitable framework for these studies. Finally, the development of machine learning-based surrogate models may offer significant acceleration of predictions while maintaining the required accuracy. In this context, graph neural network architectures appear particularly well-suited~\cite{SanchezGonzalez2020, CHOI2024, LI2024, DESHPANDE2024}.

All code, together with the experimental and simulation datasets, is made available in open-access repositories~\cite{chodkiewiczGitHub2025vpp, chodkiewiczZenodo2025vpp}. These resources are intended to facilitate reproducibility, enable further comparative studies, and serve as a foundation for future developments in both academic and industrial contexts related to particle damping, granular materials modeling, and discrete element simulations. The present work also constitutes a part of the ongoing research efforts within our group, aimed at advancing the modeling, analysis, and optimization of granular-based damping systems. We hope that by sharing these developments, we can contribute to the broader community of researchers and practitioners working in the fields of granular systems, particle mechanics, and structural dynamics.


\end{document}